\documentclass[%
 aip,jcp
 amsmath,amssymb,
 reprint,%
]{revtex4-1}
\usepackage{hyperref}
\usepackage{graphicx}
\usepackage{dcolumn}
\usepackage{bm}
\usepackage{physics}
\usepackage[utf8]{inputenc}
\usepackage[T1]{fontenc}
\usepackage{etoolbox}
\usepackage{makecell}

\makeatletter
\def\@email#1#2{%
 \endgroup
 \patchcmd{\titleblock@produce}
  {\frontmatter@RRAPformat}
  {\frontmatter@RRAPformat{\produce@RRAP{*#1\href{mailto:#2}{#2}}}\frontmatter@RRAPformat}
  {}{}
}%
\makeatother
\begin{document}

\preprint{AIP/123-QED}

\title[Improved Optimization for the Neural-network Quantum States and Tests on the Chromium Dimer]{Improved Optimization for the Neural-network Quantum States and Tests on the Chromium Dimer}
\author{Xiang Li}
\thanks{Author to whom correspondence should be addressed:\\
\href{mailto:xiangxmi6@gmail.com}{xiangxmi6@gmail.com}\\
\href{mailto:hshu@mail.tsinghua.edu.cn}{hshu@mail.tsinghua.edu.cn}
}
\affiliation{ 
Department of Chemistry and Engineering Research Center of Advanced Rare-Earth Materials\\ of Ministry of Education, Tsinghua University, Beijing 100084, China.
}
\author{Jia-Cheng Huang}
\affiliation{ 
Department of Chemistry and Engineering Research Center of Advanced Rare-Earth Materials\\ of Ministry of Education, Tsinghua University, Beijing 100084, China.
}
\author{Guang-Ze Zhang}
\affiliation{ 
Department of Chemistry and Engineering Research Center of Advanced Rare-Earth Materials\\ of Ministry of Education, Tsinghua University, Beijing 100084, China.
}
\author{Hao-En Li}
\affiliation{ 
Department of Chemistry and Engineering Research Center of Advanced Rare-Earth Materials\\ of Ministry of Education, Tsinghua University, Beijing 100084, China.
}
\author{Zhu-Ping Shen}
\affiliation{ 
Department of Chemistry and Engineering Research Center of Advanced Rare-Earth Materials\\ of Ministry of Education, Tsinghua University, Beijing 100084, China.
}
\author{Chen Zhao}
\affiliation{ 
Department of Chemistry and Engineering Research Center of Advanced Rare-Earth Materials\\ of Ministry of Education, Tsinghua University, Beijing 100084, China.
}
\author{Jun Li}
\affiliation{ 
Department of Chemistry and Engineering Research Center of Advanced Rare-Earth Materials\\ of Ministry of Education, Tsinghua University, Beijing 100084, China.
}
\affiliation{Department of Chemistry and Guangdong Provincial Key Laboratory of Catalytic Chemistry, Southern University of Science and Technology, Shenzhen 518055, China.}
\affiliation{Fundamental Science Center of Rare Earths, Ganjiang Innovation Academy, Chinese Academy of Sciences, Ganzhou 341000, China.}
\author{Han-Shi Hu}
\thanks{Author to whom correspondence should be addressed:\\
\href{mailto:xiangxmi6@gmail.com}{xiangxmi6@gmail.com}\\
\href{mailto:hshu@mail.tsinghua.edu.cn}{hshu@mail.tsinghua.edu.cn}
}
\affiliation{ 
Department of Chemistry and Engineering Research Center of Advanced Rare-Earth Materials\\ of Ministry of Education, Tsinghua University, Beijing 100084, China.
}

\date{\today}

\begin{abstract}
The advent of Neural-network Quantum States (NQS) has significantly advanced wave function ansatz research, sparking a resurgence in orbital space variational Monte Carlo (VMC) exploration. This work introduces three algorithmic enhancements to reduce computational demands of VMC optimization using NQS: an adaptive learning rate algorithm, constrained optimization, and block optimization. We evaluate the refined algorithm on complex multireference bond stretches of $\rm H_2O$ and $\rm N_2$ within the cc-pVDZ basis set and calculate the ground-state energy of the strongly correlated chromium dimer ($\rm Cr_2$) in the Ahlrichs SV basis set. Our results achieve superior accuracy compared to coupled cluster theory at a relatively modest CPU cost. This work demonstrates how to enhance optimization efficiency and robustness using these strategies, opening a new path to optimize large-scale Restricted Boltzmann Machine (RBM)-based NQS more effectively and marking a substantial advancement in NQS's practical quantum chemistry applications.
\end{abstract}

\maketitle

%

\section{\label{sec:Introduction}Introduction}

Solving the many-electron Schrödinger equation is a pivotal challenge within electronic structure theory, offering profound insights into all molecular properties. The inherent complexity arises from electron correlation, overlooked in mean-field approximations and categorized into dynamic and static correlations. Dynamic correlation, originating from short-range instantaneous Coulomb repulsion among electrons, predominates in systems where a single Slater determinant suffices for a qualitatively correct description. Coupled cluster (CC) theory\cite{CCSD82,CCSD(T)89} stands out as one of the most successful methods for introducing excitation configurations to amend these single-reference wave functions. However, in systems with stretched covalent bonds and metal-metal multiple bonds, where degenerate or quasi-degenerate electronic states make multiple determinants significant, single-reference methods fall short. Static correlation assumes paramount importance in these scenarios, with complete active space self-consistent field (CASSCF)\cite{CASSCF80} as the standard method for a qualitatively reasonable reference. Enhanced accuracy can subsequently be achieved by incorporating remaining dynamic correlation through multireference perturbation theory (MRPT),\cite{MRPT90,MRPT01,MRPT09,MRPT14_liu,MRPT14_chen} multireference configuration interaction (MRCI) methods,\cite{MRCI78,MRCI88,MRCI11} or multireference coupled cluster (MRCC) theories.\cite{MRCC12,MRCC18} Full configuration interaction (FCI) method, though capable of capturing both types of electron correlation within complete active spaces, is confined to small systems due to computational demands. Recent advances in approximate FCI solvers such as Density Matrix Renormalization Group (DMRG),\cite{DMRG11,DMRG20} Full Configuration Interaction Quantum Monte Carlo (FCIQMC),\cite{FCIQMC09,iFCIQMC10,sFCIQMC12,FCIQMC12-F12,FCIQMC14,FCIQMC19} and Selected Configuration Interaction (SCI) methods\cite{iCI16,iCI20,iCI21,ASCI16,ASCI20,ASCI-SCF20,HCI16,HCI17,HCI18,HCI20,HCI23,HCISCF17,HCISCF21} have notably enhanced the efficiency in accurately describing electron correlation within large active spaces. However, precise computations of large, complex molecular systems demand further exploration.

Neural-network Quantum States (NQS)\cite{NQS17} have emerged as a promising avenue, utilizing artificial neural networks to encapsulate the many-body wave function in a compact, second-quantized formulation. This innovative ansatz has shown remarkable success\cite{NQS23-review} in molecular\cite{NQS20-nc,NQS20-jctc,NAQS22,MADE23,NQS23-SC,NNQS-transformer,NQS24-backflow} and solid-state\cite{NQS21-solid} systems, and has been extended to both continuous spaces\cite{PauliNet20,FermiNet20,PsiFormer22,FermiNet22-ECP,FermiNet23-DMC,LapNet24} and quantum computing.\cite{QML2018,QCNQS2023,QST2018,QCNQS2020,QCQNS2022} The energy evaluation and NQS ansatz optimization usually require the Orbital Space Variational Monte Carlo (VMC) framework, whose typical complexity is $O(N^5)$ but can be improved to $O(N^4)$\cite{VMC18-N^3} or lower\cite{VMC18-N^2} with strategic approximations, where $N$ is the system size. Nonetheless, the computational overhead introduced by neural networks, compared to traditional wave function ansatzes, has limited NQS's broader applications in molecules. A notable challenge lies in the inefficiency of Monte Carlo sampling within the VMC framework. One of the promising solutions is exact sampling via autoregressive neural networks,\cite{NAQS22} as demonstrated with the weakly correlated $\rm Na_2CO_3$ in minimal basis sets, showcasing competitive accuracy to the Coupled Cluster with Singles and Doubles (CCSD) method.\cite{MADE23} More recently, we introduced another deterministic configuration selection scheme to avoid Monte Carlo sampling,\cite{NQS23-SC} eliminating stochastic noise and enhancing the applicability of various neural network types.\cite{NQS24-backflow} Building on this insight, in this article, we detail three algorithmic advancements to optimize NQS through orbital space VMC calculations and examine the performance of NQS predicated on Restricted Boltzmann Machine (RBM) model in tackling large, strongly correlated molecular systems.

The rest of this article unfolds as follows: Sec. \ref{sec:overview} offers a succinct overview of the VMC framework. Subsequently, Sec. \ref{sec:adapt} introduces an adaptive learning rate algorithm. Sec. \ref{sec:constrained} elaborates on a constrained optimization strategy within a selected configuration subspace. Then Sec. \ref{sec:block} outlines a divide-and-conquer strategy tailored for large RBM-based NQS within the stochastic reconfiguration (SR) method---block optimization. The performance of these algorithmic enhancements is presented in sections \ref{sec:enhanced}, \ref{sec:perform_constrained}, and \ref{sec:perform_block}. We further evaluate the accuracy of RBM-based NQS in large, strongly correlated systems, including dissociation curves of $\rm H_2O$ and $\rm N_2$ in the cc-pVDZ basis set (Sec. \ref{sec:h2o_n2}), and the ground-state energy of the chromium dimer in the Ahlrichs SV basis set (Sec. \ref{sec:cr2}). Finally, we give a conclusion in Sec. \ref{sec:conclusion}.

\section{\label{sec:theory}Theory and Methods}

\subsection{\label{sec:overview}Overview}
Neural-network quantum states offer a novel approach to describing the many-electron wave function in Hilbert space. Utilizing a flexible neural network model, the NQS wave function ansatz is formulated as
\begin{equation}
|\mathrm{\Psi}_\theta\rangle=\sum_{k}{\hat{P}\psi_\theta\left(\sigma_1^k,\ \sigma_2^k,\ \ldots,\ \sigma_M^k\right)|D_k\rangle},
\end{equation}
where $|D_k\rangle=|\sigma_1^k,\ \sigma_2^k,\ \ldots,\ \sigma_M^k\rangle$ represents an electron configuration in Slater determinant form. Here, $M$ denotes the total number of spin orbitals, and $\sigma_i^k\in\left\{0,1\right\}$ indicates the occupation number of each spin orbital. To maintain symmetry, a projector $\hat{P}$, is employed to projects out the electron configurations that satisfy the point group symmetry.\cite{NQS18-symm} Additionally, spin symmetry is incorporated through a preprocessing step of the network input.\cite{NAQS22}

In this study, we employ the restricted Boltzmann machine model, and define the coefficient function $\psi_\theta$ as
\begin{equation}\label{eq:RBM_psi}
\psi_\theta\left(\sigma_1^k,\sigma_2^k,\ldots,\sigma_M^k\right)=e^{\sum_{i}^{M}{a_i\sigma_i^k}}\prod_{j}^{\alpha \times M}\left(1+e^{b_j+\sum_{i}^{M}{W_{ij}\sigma_i^k}}\right).
\end{equation}
This formula is particularly justified for molecular systems. The hidden unit density, $\alpha$, which is the ratio of the number of hidden units to the number of input units, determines the expressive power of RBM-based NQS. The network parameters $\theta$, comprising complex-valued input biases $a_i$, hidden biases $h_i$ and kernel weights $W_{ij}$, are typically initialized randomly. In this work, we initialize the input biases $a_i$ with the negative value of the corresponding orbital energy $\epsilon_i$, leading to the Boltzmann distribution expression
\begin{equation}\label{eq:Boltzmann-Distribution}
p_{\rm B}\left(\sigma_1^k,\sigma_2^k,\ldots,\sigma_M^k\right)=e^{\sum_{i}^{M}{a_i\sigma_i^k}}=e^{\sum_{i}^{M}-\frac{\epsilon_i\sigma_i^k}{T}},
\end{equation}
where $T=1$ is an hyperparameter, and $\epsilon_i$ (or $a_i$) are updated during NQS optimization. The Boltzmann distribution can be considered as an independent electron approximation, applicable to various NQS models, as detailed in Appendix \ref{sec:MO_Initialization}.

For optimizing the NQS wave function ansatz, we employ the orbital space VMC framework. The energy of the NQS is expressed as
\begin{equation}\label{eq:accu_energy}
E_\theta=\frac{\mel{\Psi_\theta}{\hat{H}}{\Psi_\theta}}{\langle\mathrm{\Psi}_\theta|\mathrm{\Psi}_\theta\rangle}=\sum_{k}{P\left(D_k\right)E_{\rm loc}}\left(D_k\right),\ \ 
\end{equation}
where $P\left(D_k\right)$ and $E_{\rm loc}\left(D_k\right)$ represent the probability and local energy of a given electron configuration, respectively. The probability $P\left(D_k\right)$ is defined as
\begin{equation}\label{eq:accu_prob}
P\left(D_k\right)=\frac{\left|\langle D_k|\mathrm{\Psi}_\theta\rangle\right|^2}{{\sum_{i}\left|\langle D_i|\mathrm{\Psi}_\theta\rangle\right|}^2},
\end{equation}
and the local energy $E_{\rm loc}\left(D_k\right)$ is given by
\begin{equation}\label{eq:accu_loc}
E_{\rm loc}\left(D_k\right)=\frac{\mel{D_k}{\hat{H}}{\Psi_\theta}}{\langle D_k|\mathrm{\Psi}_\theta\rangle}=\sum_{\mu}\frac{\langle D_\mu|\mathrm{\Psi}_\theta\rangle}{\langle D_k|\mathrm{\Psi}_\theta\rangle}
\langle{D_k}|{\hat{H}}|{D_\mu}\rangle.
\end{equation}
This intractable energy expression can be approximated by stochastically sampling a set of electron configurations, $\mathcal{V}$, from the probability distribution $P\left(D_k\right)$. An alternative, more efficient nonstochastic strategy, as detailed in our previous work,\cite{NQS23-SC} builds such a sample $\mathcal{V}$ by selecting configurations with significant amplitude coefficients when calculating local energies. Employing this deterministically selected configuration sample $\mathcal{V}$, the energy value of NQS is estimated by
\begin{equation}\label{eq:appr_energy}
E_\theta\approx\sum_{k\in\mathcal{V}}{P\left(D_k\right)E_{\rm loc}}\left(D_k\right),
\end{equation}
\begin{equation}\label{eq:appr_prob}
P\left(D_k\right)\approx\frac{\left|\langle D_k|\mathrm{\Psi}_\theta\rangle\right|^2}{{\sum_{i\in\mathcal{V}}\left|\langle D_i|\mathrm{\Psi}_\theta\rangle\right|}^2}.
\end{equation}

Following the variational principle, the NQS ansatz can progressively approximate the exact ground-state wave function by iteratively minimizing its energy expectation value. This is achieved by adjusting the network parameters $\mathcal{\theta}$ using gradient-based optimization methods. In the VMC framework, the gradient is approximated by
\begin{equation}\label{eq:grad}
g_{\theta_m}=\frac{\partial E_\theta}{\partial\theta_m^\ast}\approx\sum_{k\in\mathcal{V}}{P\left(D_k\right)O_{{k\theta}_m}^\ast\left[E_{\rm loc}\left(D_k\right)-E_\theta\right]},
\end{equation}
where $O_{k\theta_m}$ represents the gradient of the coefficient function, {$\ln\psi_\theta$}, with respect to network parameters. We employ the SR scheme\cite{SR01,SR07} to treat the gradients and update the network parameters. The parameter update direction, $\mathrm{\Delta\theta}$, is determined through the linear equation
\begin{equation}\label{eq:SR}
S\Delta\theta=g_\theta,
\end{equation}
with
\begin{equation}
\begin{aligned}
    S_{mn}\approx&\sum_{k\in\mathcal{V}}{P\left(D_k\right)O_{k\theta_m}^\ast O_{k\theta_n}}-\\&\sum_{k\in\mathcal{V}}{P\left(D_k\right)O_{k\theta_m}^\ast}\sum_{k\in\mathcal{V}}{P\left(D_k\right)O_{k\theta_n}}+\lambda\delta_{mn}.
\end{aligned}
\end{equation}
In this context, $\lambda $ is a regularization parameter that stabilizes the optimization process.\cite{SR07} In our experiments, smaller values of $\lambda$ generally lead to higher accuracy, provided the optimization remains stable. After solving this linear equation for $\Delta \theta$, parameters are updated as follows
\begin{equation}\label{eq:update}
\theta^\prime=\theta-\eta\Delta\theta,
\end{equation}
where $\eta$ is the step size or learning rate.

\subsection{\label{sec:adapt}Adaptive learning rate}
In the previously discussed VMC framework, the learning rate hyperparameter, $\eta$, plays a crucial role in optimization performance. Commonly, a constant, small value of $\eta$ is employed to ensure robustness, albeit at the expense of slower convergence. The challenge of selecting an optimal learning rate can be mitigated by using finely-tuned adaptive optimizers like RMSprop,\cite{RMSProp17} AMSGrad\cite{VMC18-N^2,AMSGrad19} and ADAM,\cite{Adam14} which may be further integrated with learning rate schedulers in TensorFlow\cite{TensorFlow} or PyTorch.\cite{PyTorch} However, the initial learning rate still significantly impacts the quality of optimization.\cite{VMC19-ParamsOpt} Another adaptive strategy within the SR scheme involves selecting a learning rate that minimizes the energy expectation value within a specified range at each iteration.\cite{adap_lr21} This approach promotes robust convergence patterns and mirrors the linear method (LM) in searching for an optimal shift value of matrix diagonals,\cite{adpt_diag_LM07} preventing excessively large steps in parameter space. Nonetheless, evaluating energy expectation values across all potential learning rates significantly increases computational time, thus limiting the number of feasible candidates.

To streamline this process, we propose a method to efficiently estimate the energy values of multiple wave function ansatzes obtained with varying learning rates. At each iteration, we first assess the overlap between the original NQS ansatz, $\mathrm{\Psi}_\theta$, and all updated ansatzes, $\mathrm{\Psi}_{\theta^\prime}$, for different learning rates within the range of $\rm {lr}_{min}\leq\eta\leq{lr}_{max}$. The overlap is calculated as follows:
\begin{equation}
\sqrt{\frac{\langle\mathrm{\Psi}_\theta|\mathrm{\Psi}_{\theta^\prime}\rangle\langle\mathrm{\Psi}_{\theta^\prime}|\mathrm{\Psi}_\theta\rangle}{\langle\mathrm{\Psi}_\theta|\mathrm{\Psi}_\theta\rangle\langle\mathrm{\Psi}_{\theta^\prime}|\mathrm{\Psi}_{\theta^\prime}\rangle}}.
\end{equation}
We consider the learning rates of these updated ansatzes only if the overlap exceeds 0.98. This criterion prevents substantial transitions between $\mathrm{\Psi}_\theta$ and $\mathrm{\Psi}_{\theta^\prime}$, ensuring that the sample $\mathcal{V}$ deterministically selected during the energy calculation for $\mathrm{\Psi}_\theta$ can reliably approximate the important configuration sample for $\mathrm{\Psi}_{\theta^\prime}$.

Subsequently, we evaluate the energy of all remaining updated ansatzes, selecting the one with the minimum energy for this iteration's update. A pivotal aspect of enhancing this energy evaluation process involves truncating the NQS ansatz in the full configuration space to an approximation in the selected configuration sample $\mathcal{V}$
\begin{equation}
|\mathrm{\Psi}_\theta\rangle\approx|{\widetilde{\mathrm{\Psi}}}_\theta\rangle \equiv\sum_{k\in\mathcal{V}}{\hat{P}\psi_\theta\left(\sigma_1^k,\ \sigma_2^k,\ \ldots,\ \sigma_M^k\right)|D_k\rangle}.
\end{equation}
An exact variational energy expression for this truncated NQS ansatz is
\begin{equation}\label{eq:truncated_energy}
{\widetilde{E}}_\theta=\frac{\langle \widetilde{\Psi}_\theta|{\hat{H}}|{\widetilde{\Psi}_\theta}\rangle}{\langle{\widetilde{\mathrm{\Psi}}}_\theta|{\widetilde{\mathrm{\Psi}}}_\theta\rangle}=\sum_{k\in\mathcal{V}}{\widetilde{P}\left(D_k\right){\widetilde{E}}_{\rm loc}}\left(D_k\right),
\end{equation}
with the modified probability and local energy being
\begin{equation}\label{eq:truncated_prob}
\widetilde{P}\left(D_k\right)=\frac{\left|\langle D_k|\mathrm{\Psi}_\theta\rangle\right|^2}{{\sum_{i\in\mathcal{V}}\left|\langle D_i|\mathrm{\Psi}_\theta\rangle\right|}^2},
\end{equation}
\begin{equation}\label{eq:truncated_loc}
{\widetilde{E}}_{\rm loc}(D_k)=\frac{\langle D_k|\hat{H}|{\widetilde{\mathrm{\Psi}}}_\theta\rangle}{\langle D_k|{\widetilde{\mathrm{\Psi}}}_\theta\rangle}=\sum_{\mu\in\mathcal{V}}\frac{\langle D_\mu|{\widetilde{\mathrm{\Psi}}}_\theta\rangle}{\langle D_k|{\widetilde{\mathrm{\Psi}}}_\theta\rangle}\mel{D_k}{\hat{H}}{D_\mu}.
\end{equation}

This truncated energy expression $\widetilde{E}_\theta$ is employed to select the optimal learning rate and significantly reduces the requisite evaluations for $\psi_\theta$, which are computationally intensive due to the complexity of neural networks. The computational demand for calculating the coefficient function $\psi_\theta$ under Eq. (\ref{eq:accu_loc}) incurs a complexity of  $O\left(N_\mathcal{V}\times N_{\rm sd}\times N_{\rm wf}\right)$. Here, $N_\mathcal{V}$ represents the size of the selected configuration sample $\mathcal{V}$, $N_{\rm sd}$ denotes the size of the single and double excitation space for each electron configuration, and $N_{\rm wf}$ is the number of updated NQS ansatzes generated using all potential learning rates. However, this complexity is reduced to $O\left(N_\mathcal{V}\times N_{\rm wf}\right)$ when employing Eq. (\ref{eq:truncated_loc}), since these calculations solely involve configurations within the selected configuration sample $\mathcal{V}$. Additionally, the computational effort required for evaluating Hamiltonian matrix elements shifts from  $O\left(N_\mathcal{V}\times N_{\rm sd}\right)$ to competitive $O\left(N_\mathcal{V}\times N_\mathcal{V}\right)$. Therefore, the number of potential learning rates, $N_{\rm wf}$, no longer significantly impacts computational expense as before.

\subsection{\label{sec:constrained}Constrained optimization within subspace}

Within the VMC framework, optimizing the NQS ansatz primarily focuses on minimizing its energy expectation value. By employing the adaptive learning rate strategy, a truncated energy value, ${\widetilde{E}}_\theta$, is obtained through Eq. (\ref{eq:truncated_energy}), providing an alternative path to optimize network parameters via its gradient
\begin{equation}
{\widetilde{g}}_{\theta_m}=\frac{\partial{\widetilde{E}}_\theta}{\partial\theta_m^\ast}=\sum_{k\in\mathcal{V}}{\widetilde{P}\left(D_k\right)O_{{k\theta}_m}^\ast\left[{\widetilde{E}}_{\rm loc}\left(D_k\right)-{\widetilde{E}}_\theta\right]}.
\end{equation}
We propose a constrained optimization workflow tailored to the subspace defined by the selected configuration sample $\mathcal{V}$, updating network parameters as follows:

\begin{enumerate}
\item Deterministically update the configuration sample $\mathcal{V}$ during the evaluation of energy $E_\theta$ in Eq. (\ref{eq:accu_energy});
\item Calculate the energy gradient $g_\theta$ of $E_\theta$ and apply the adaptive learning rate algorithm to determine parameter updates, thus generating the truncated energy value ${\widetilde{E}}_\theta$ for the updated NQS ansatz;
\item Evaluate the gradient ${\widetilde{g}}_\theta$ of ${\widetilde{E}}_\theta$ using the same configuration sample $\mathcal{V}$, and reapply the adaptive learning rate algorithm for parameter updates;
\item Repeat Step 3 as desired while keeping the sample $\mathcal{V}$ from step 2 unchanged;
\item Proceed to the next iteration.
\end{enumerate}

\begin{figure}[!h]
\includegraphics[width=0.45\textwidth]{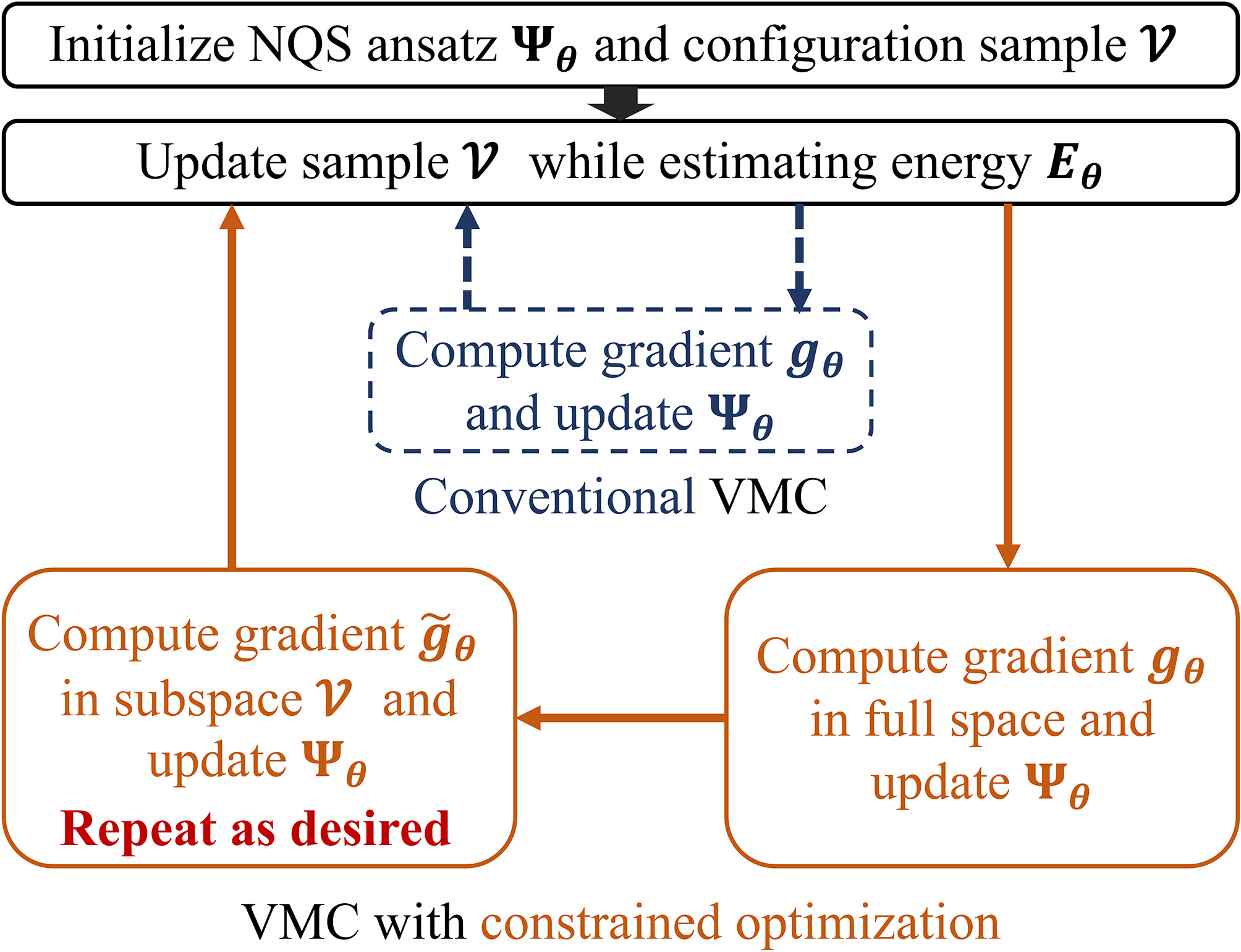}
\caption{\label{fig:flowchart} The flowchart illustrating the comparison between the conventional VMC framework and the VMC algorithm enhanced with constrained optimization.}
\end{figure}

This iterative process ensures $N_{\rm opt}$ updates of the NQS ansatz within a single iteration, including one update utilizing $E_\theta$ and $N_{\rm opt}-1$ subsequent updates using ${\widetilde{E}}_\theta$. To facilitate comprehension, a schematic illustrating this optimization process is included in Figure \ref{fig:flowchart}.

The effectiveness of this constrained optimization stems from the minimal changes to the NQS ansatz $\mathrm{\Psi}_\theta$ after $N_{\rm opt}$ parameter updates, with the selected configuration sample $\mathcal{V}$ serving as an accurate approximation for the updated ansatz $\mathrm{\Psi}_{\theta^\prime}$. The requisite overlap between the original and final ansatzes exceeds 0.98, ensuring accuracy. Additionally, the truncated energy value, ${\widetilde{E}}_\theta$, assessed during the adaptive learning rate algorithm application, is then reused for updating network parameters with its gradient, ${\widetilde{g}}_\theta$. As $N_{\rm opt}$ increases, the cost of this constrained optimization diminishes, benefiting from the repeated utilization of ${\widetilde{E}}_\theta$. Moreover, the initial update using $E_\theta$ functions as a configuration selection mechanism, while the subsequent $N_{\rm opt}-1$ updates leveraging ${\widetilde{E}}_\theta$ resemble a rough diagonalization of the Hamiltonian matrix within the subspace of the selected configuration sample $\mathcal{V}$. Therefore, the overall NQS ansatz optimization process parallels SCI methods.

\subsection{\label{sec:block}Block optimization for the restricted Boltzmann machine}

The SR method has demonstrated high accuracy and robustness in optimizing NQS. However, the optimization of the NQS ansatz for large molecular systems using the SR method poses a significant challenge due to the expansive size of the matrix $S$ in Eq. (\ref{eq:SR}). This complexity is mitigated in other deep neural-network models, such as FermiNet,\cite{FermiNet20} through the adoption of a modified Kronecker-factored approximate curvature (KFAC) method.\cite{KFAC15} The KFAC method simplifies optimization by segmenting parameters into blocks by layers and approximating the $S$ matrix as block diagonal. 

In this work, we discover that the parameters of the RBM model could also be divided into blocks, exploiting the property that the analytical form of an RBM model can be segmented into a product of several smaller RBM models. Specifically, the coefficient function $\psi_{\rm RBM}$ can be decomposed as

\begin{equation}
\begin{aligned}
\psi_{\rm RBM}&=e^{\sum_{i}^{M}{a_i\sigma_i^k}}\prod_{j}^{\alpha\times M}\left(1+e^{b_j+\sum_{i}^{M}{W_{ij}\sigma_i^k}}\right)\\
&=\prod_{q}^{N_b}\left\{e^{\sum_{i}^{M}\frac{a_i\sigma_i^k}{N_b}}\prod_{j_q}^{\frac{\alpha\times M}{N_b}}\left(1+e^{b_{j_q}+\sum_{i}^{M}{W_{ij_q}\sigma_i^k}}\right)\right\}\\
&=\prod_{q}^{N_b}\psi_{\rm rbm}^{\left(q\right)}.
\end{aligned}
\end{equation}

\begin{figure}[!h]
\includegraphics[width=0.4\textwidth]{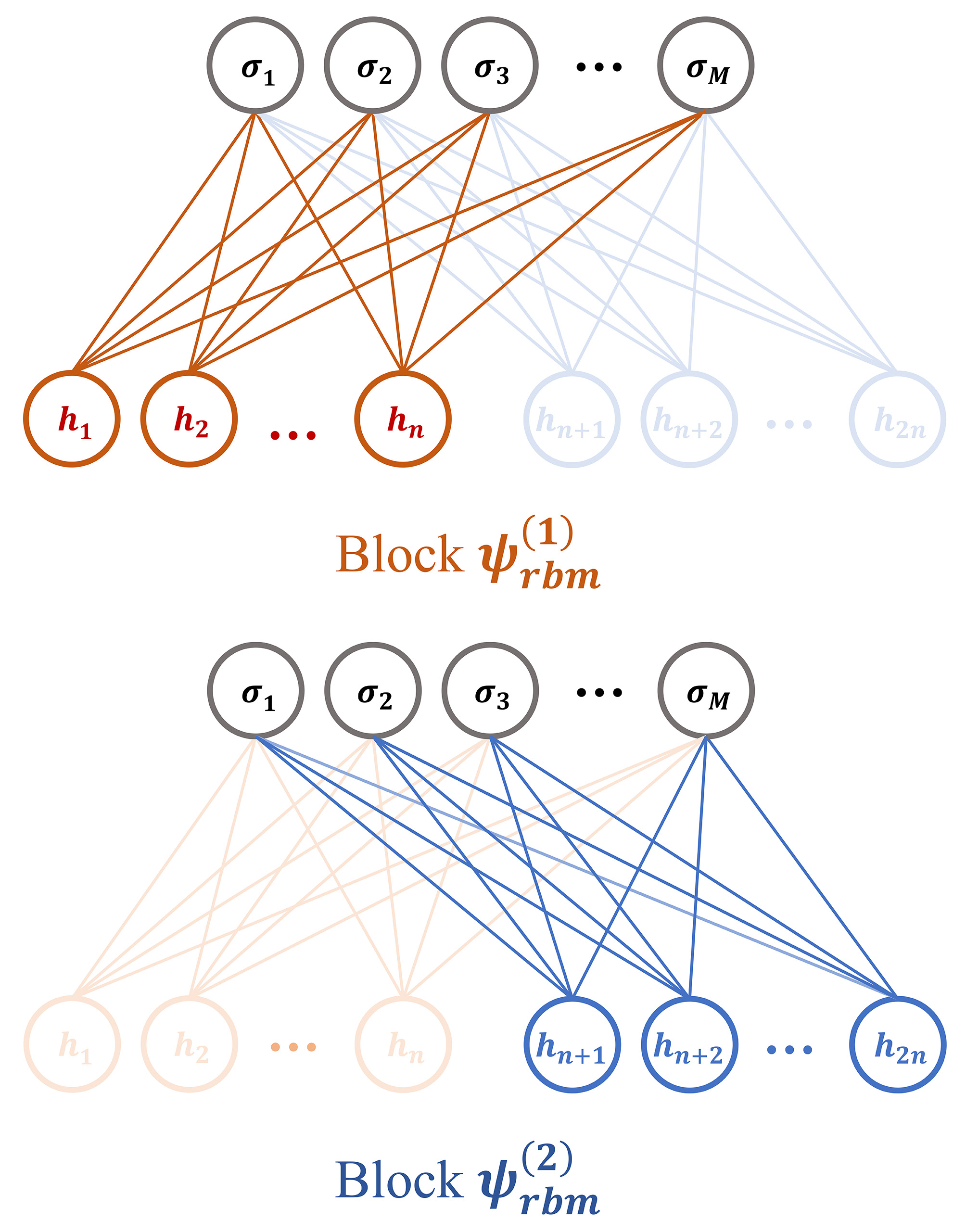}
\caption{\label{fig:rbm} The figure showcases the partitioning of one RBM model with $2n$ hidden units into two blocks: the orange part and the blue part, each containing $n$ hidden units. }
\end{figure}

This decomposition illustrates how the SR method can be applied to each block individually, thus facilitating the optimization of significantly large RBM models. The strategy of dividing parameters into blocks can be static, remaining unchanged throughout the optimization process, or dynamic, with blocks being re-divided at each parameter update. Furthermore, our experiments have demonstrated the efficacy of optimizing a subset of the blocked parameters at a time. To aid visualization, we include a graphic representation (Figure \ref{fig:rbm}) of the RBM division when $N_b=2$, underscoring the practical implementation of this block optimization strategy. We note that Zhang \textit{et al.}\cite{block_RBM23} have recently also proposed one sequential local optimization strategy for RBM, which exploits the locality of the Hamiltonian and a segmentation of the input units.

\section{\label{sec:Results}Results and Discussion}

We initiate our discussion by examining the symmetric stretching potential energy curve of $\rm H_2O$ using the 6-31G basis set. This analysis aims to evaluate the performance of three key methodologies: the adaptive learning rate algorithm, constrained optimization within the subspace, and block optimization for the RBM model. For these calculations, we deal with the system's full configuration space, which adheres to $C_{2v}$ symmetry and encompasses approximately $4.1\times{10}^5$ electron configurations. During our analysis, the angle between the two O$-$H bonds is held constant at 104.2°, across 20 distinct O$-$H bond lengths ranging from 0.8 Å to 4 Å. 

Subsequently, we extend our investigation to the dissociation curves for $\rm H_2O$ and $\rm N_2$ in the cc-pVDZ basis set, incorporating the aforementioned three methodological enhancements. The structural parameters for $\rm H_2O$ are kept consistent, and we explore 17 different N$-$N bond lengths, from 0.8 Å to 2.8 Å. The full configuration spaces for $\rm H_2O$ and $\rm N_2$ (with two frozen orbitals) are roughly $4.5\times{10}^8$ and $5.4\times{10}^{8}$ electron configurations, respectively, maintaining $C_{2v}$ symmetry for $\rm H_2O$ and $ D_{2h}$ symmetry for $\rm N_2$. Additionally, we conduct a preliminary investigation of the chromium dimer ($\rm Cr_2$) at a bond length of 1.5 Å using the Ahlrichs SV basis.\cite{AhlrichsSV92} Within its active space (24e,30o), the total possible electron configurations with $ D_{2h}$ symmetry approximate $9.4\times{10}^{14}$, escalating to an immense $1.6\times{10}^{22}$ in its full configuration space (48e,42o).

The FCI results and all orbital integrals are generated using the PySCF packages.\cite{PySCF18,PySCF20} One- and two-electron integrals of the Hamiltonian are screened with a threshold of ${10}^{-6}$ when calculating the local energy in Eq. (\ref{eq:accu_loc}).\cite{VMC18-N^2} The hidden unit density, $\alpha$, is set at 2 for calculations in the 6-31G basis and at 6 in the cc-pVDZ basis, unless specified otherwise. The regularization parameter, $\lambda$, is ${10}^{-4}$ in the 6-31G basis and ${10}^{-5}$ in the cc-pVDZ basis. Convergence criteria are defined based on changes in NQS energy: less than $N_{\rm opt}\times{10}^{-7}$ Hartree (Ha) for $10$ consecutive iterations in the 6-31G basis and less than $N_{\rm opt}\times{10}^{-6}$ Ha in the cc-pVDZ basis calculation. Here, $N_{\rm opt}$ represents the number of network parameter updates, including those from constrained optimization, per iteration. Additionally, the selected configuration sample, $\mathcal{V}$, comprises electron configurations with $\left|\psi_\theta\right|>{10}^{-5}$ in the intermediate normalized form of NQS ansatz. Detailed setups for $\rm Cr_2$ calculations are discussed in the subsequent section (Sec. \ref{sec:cr2}).

\subsection{\label{sec:enhanced}Performance of adaptive learning rate}

This subsection is dedicated to evaluating the adaptive learning rate algorithm's performance relative to the constant learning rate approach. This comparison entails computing the ground-state energies of $\rm H_2O$ in the 6-31G basis set across 20 varied O$-$H bond lengths. In the constant learning rate regime, eight distinct learning rates ($\eta_0$) are investigated: 0.01, 0.02, 0.05, 0.10, 0.15, 0.20, 0.25 and 0.30. Correspondingly, the adaptive learning rate algorithm is configured to dynamically select from 100 potential $\eta$ values within the range $0.001\le\eta\le\eta_0$. The calculations persist until convergence is achieved or the iteration count reaches 10,000.

\begin{figure}[!h]
    \centering
    \includegraphics[width=0.45\textwidth]{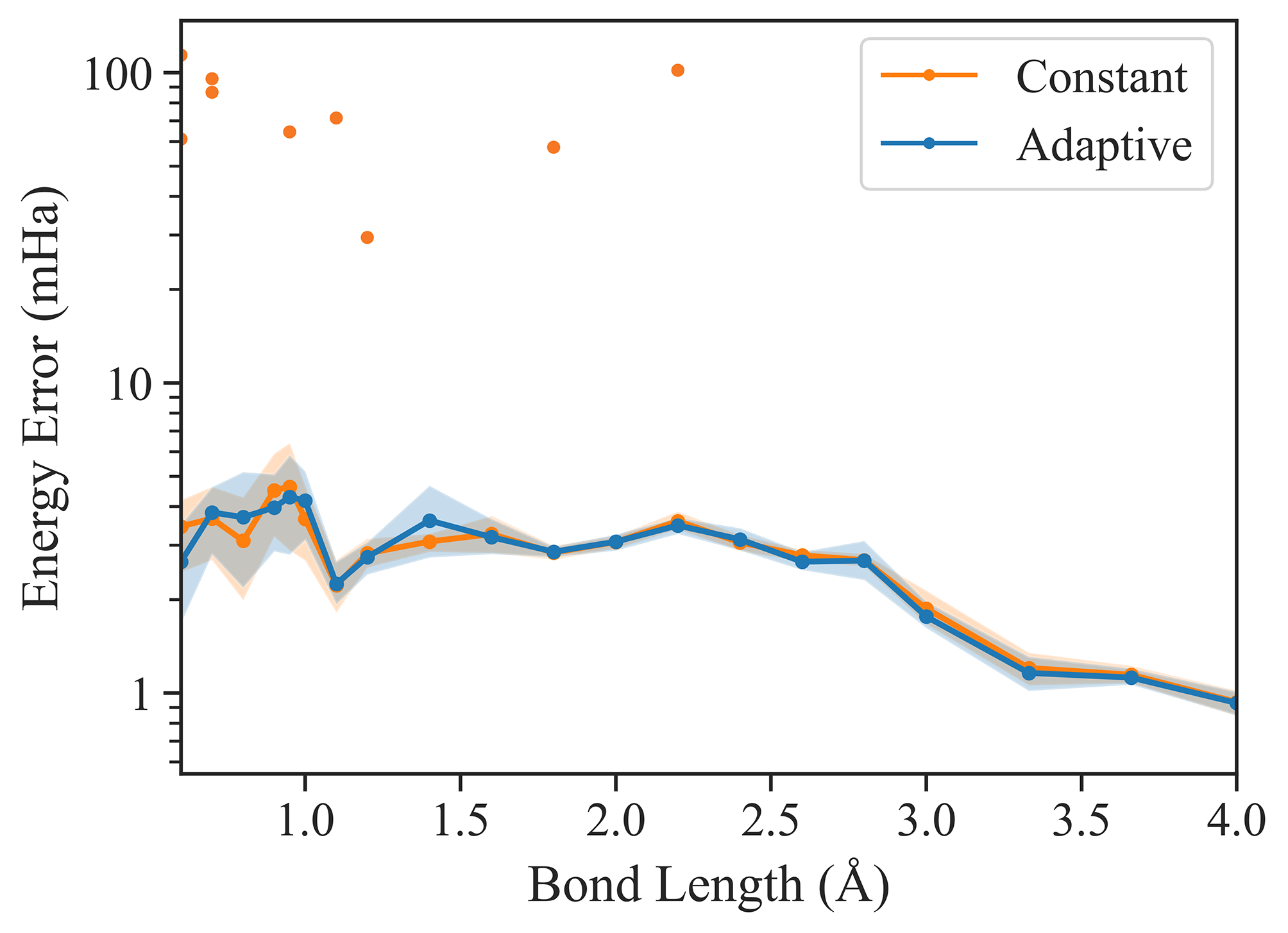}
    \caption{Energy errors relative to FCI results for ground-state energy calculations of $\rm H_2O$ across 20 O$-$H bond lengths within the 6-31G basis set, utilizing both constant and adaptive learning rate approaches. The lines represent the mean energy errors across eight different learning rate ($\eta_0$) setups, with the accompanying shaded area depicting the 95$\%$ confidence interval. Data points situated above the curve highlight instances where the use of larger constant learning rates resulted in anomalous deviations. These deviations were not taken into account in the calculation of mean energy errors.}
    \label{fig:3}
\end{figure}

\begin{figure}[!h]
    \centering
    \includegraphics[width=0.45\textwidth]{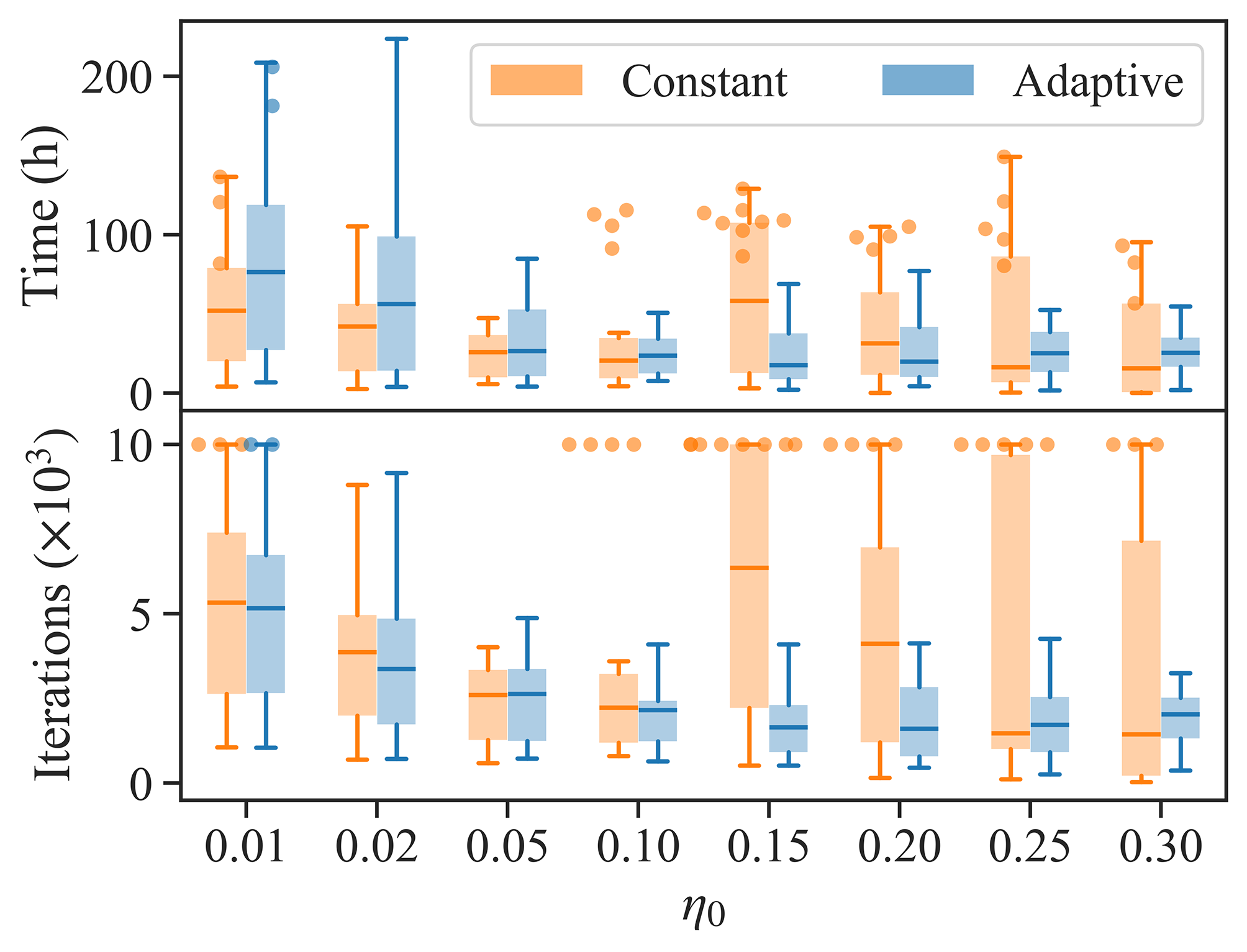}
    \caption{The comparison in computational efficiency (measured in wall time and the number of iterations) between adaptive and constant learning rate schemes employed for determining the ground-state energies of $\rm H_2O$ over 20 O$-$H bond lengths within the 6-31G basis set. The analysis spans eight distinct learning rates ($\eta_0$) values. The points signify instances where the calculations failed to converge. The rectangles show the 25-75th percentile range of the distribution, with a horizontal line marking the median. The whiskers extend to the minimum and maximum observed values.}
    \label{fig:4}
\end{figure}

As illustrated in Figure \ref{fig:3}, both adaptive and constant learning rate approaches achieve comparable levels of high accuracy. However, it is noteworthy that the relatively large constant learning rates (0.2, 0.25 and 0.30) result in some instances of falling into poor local minima, whereas the corresponding adaptive approaches do not. Regarding efficiency, as depicted in Figure \ref{fig:4}, the adaptive learning rate scheme outperforms the constant rate in terms of both wall time and iteration count to achieve similar accuracy levels when $\eta_0>0.05$. This efficiency gain stems from the adaptive method's capability to judiciously utilize small learning rates for stability and large rates for faster convergence, avoiding the oscillations near the minimum often observed with larger constant rates. This is corroborated by the observation that the large constant learning rates exhibit a higher incidence of non-convergence within 10,000 iterations. Interestingly, in cases where $\eta_0\le0.05$, the adaptive learning rate algorithm consistently selects the maximum available value ($\eta_0$), aligning with the behavior of the constant learning rate scheme in terms of iteration count. Besides,the distinctions in wall time between adaptive and constant methods reveal that the adaptive scheme incurs merely about twice the time cost per iteration due to its additional energy evaluation, a reasonable trade-off for its considerable improvements in learning rate selection at larger $\eta_0$ values.


In essence, while the constant learning rate approach demands precise configuration to ensure optimal performance, as seen with a rate of 0.05 in this study, the adaptive strategy offers a more user-friendly alternative. It seamlessly combines the advantages of both large and small learning rates---efficiency and stability, respectively---with a marginal increase in computational overhead equivalent to approximately one additional energy evaluation. This balance positions the adaptive learning rate algorithm as a superior option for seeking both robustness and enhanced performance without the intricacies of manual learning rate adjustment.

\subsection{\label{sec:perform_constrained}Performance of constrained optimization}

This segment evaluates the impact of constrained optimization on calculating the ground-state energies of $\rm H_2O$ using the 6-31G basis set, achieved by adjusting the frequency of network parameter updates per iteration ($N_{\rm opt}$). We explore a spectrum of $ N_{\rm opt}$ values: 1, 2, 3, 4, 5, 10, 20, 30, 40 and 50, where $N_{\rm opt}=1$ equates to a conventional VMC iteration augmented with our adaptive learning rate algorithm. Values of $N_{\rm  opt}$ greater than 1 incorporate $N_{\rm opt}-1$ rounds of constrained optimization within each VMC iteration.

\begin{figure}[!h]
    \centering
    \includegraphics[width=0.45\textwidth]{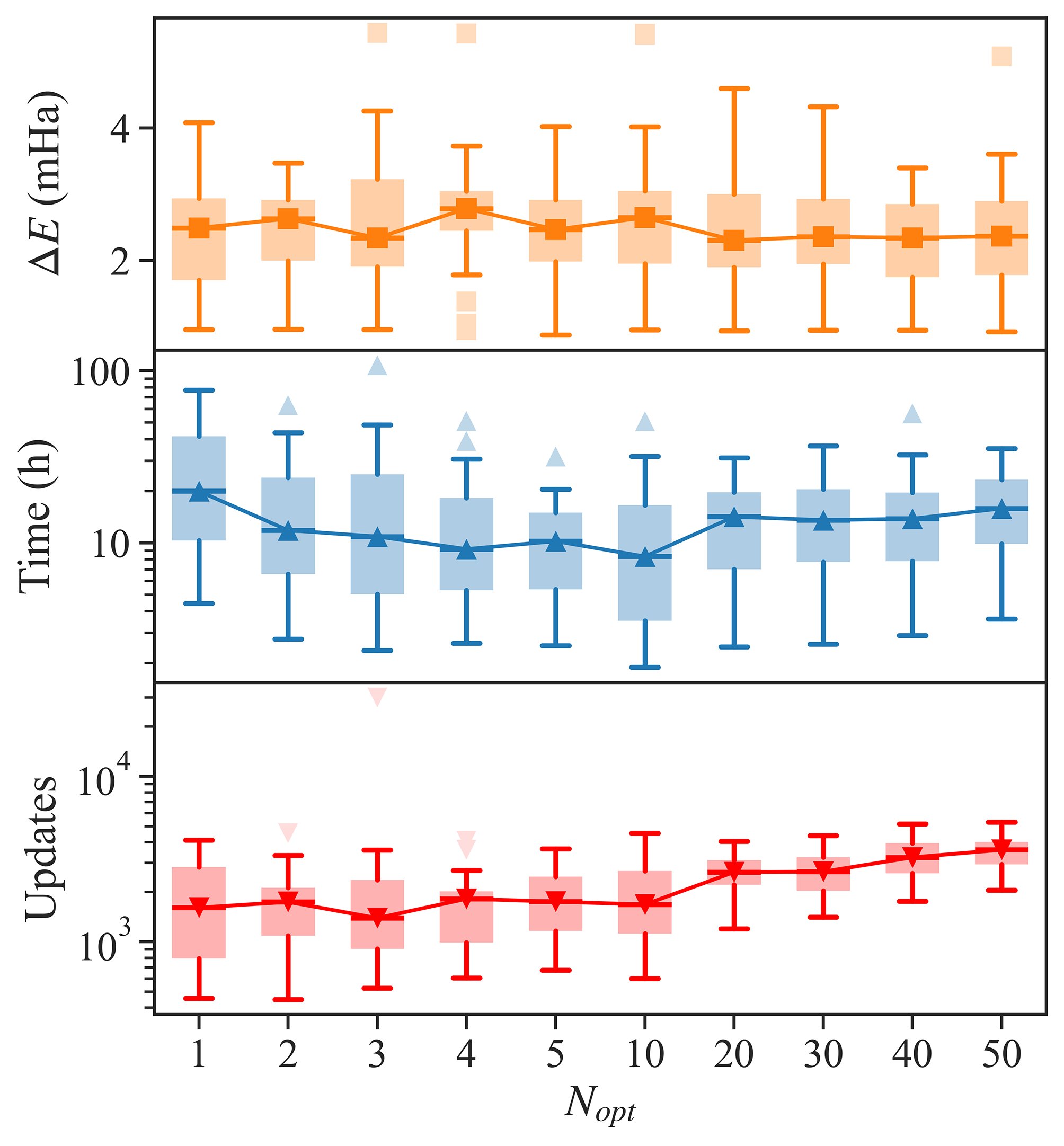}
    \caption{Performances of the constrained optimization scheme on calculating the ground-state energies of $\rm H_2O$ over 20 O$-$H bond lengths within the 6-31G basis set, examining the impact of varying the number of network parameter updates per iteration ($N_{\rm opt}$) from 1 to 50. The figure is divided into three panels for a detailed analysis: the top panel focuses on the energy differences ($\Delta E$) relative to FCI results, the middle panel showcases the required wall time, and the bottom panel illustrates the total number of parameter updates needed. Time and Updates are presented in logarithmic scale. The rectangles in each panel outline the 25th to 75th percentile of the distribution, with a horizontal line marking the median value. Whiskers extend to cover the non-outlier range, defined as 1.5 interquartile ranges above and below the boxes, while individual points outside this range are plotted separately. }
    \label{fig:5}
\end{figure}

As illustrated at the top of Figure \ref{fig:5}, all setups achieve similarly high accuracy, highlighting the effectiveness of constrained optimization. Moreover, for $N_{\rm opt}\le10$, the bottom of Figure \ref{fig:5} demonstrates a consistent number of network parameter updates across the board, suggesting that the energy ${\widetilde{E}}_\theta$, employed in constrained optimization, serves as an effective target for minimization, comparable to the conventional energy $E_\theta$ in VMC optimization. Notably, as depicted in the middle of Figure \ref{fig:5} (presented in logarithmic scale), enhanced computational efficiency is associated with higher $N_{\rm opt}$ values, particularly when $N_{\rm opt}\le10$, attributable to the decreased ratio of energy evaluations to parameter updates, benefiting from the reuse of ${\widetilde{E}}_\theta$. However, for $N_{\rm opt}>10$, the efficiency gains of constrained optimization diminish, because the significant increase in $N_{\rm opt}$ causes extra difficulties in convergence under current setups. 

This efficiency behavior diverges from the Importance Sampling Gradient Optimization (ISGO) algorithm,\cite{ISGO20} which improves computational speed by reusing the configuration sample and thus reducing the frequency of Monte Carlo sampling. In contrast, Our approach eliminates Monte Carlo sampling altogether through a deterministic configuration selection scheme\cite{NQS23-SC} and accelerates optimization by reusing ${\widetilde{E}}_\theta$.

\begin{figure}[!h]
    \centering
    \includegraphics[width=0.45\textwidth]{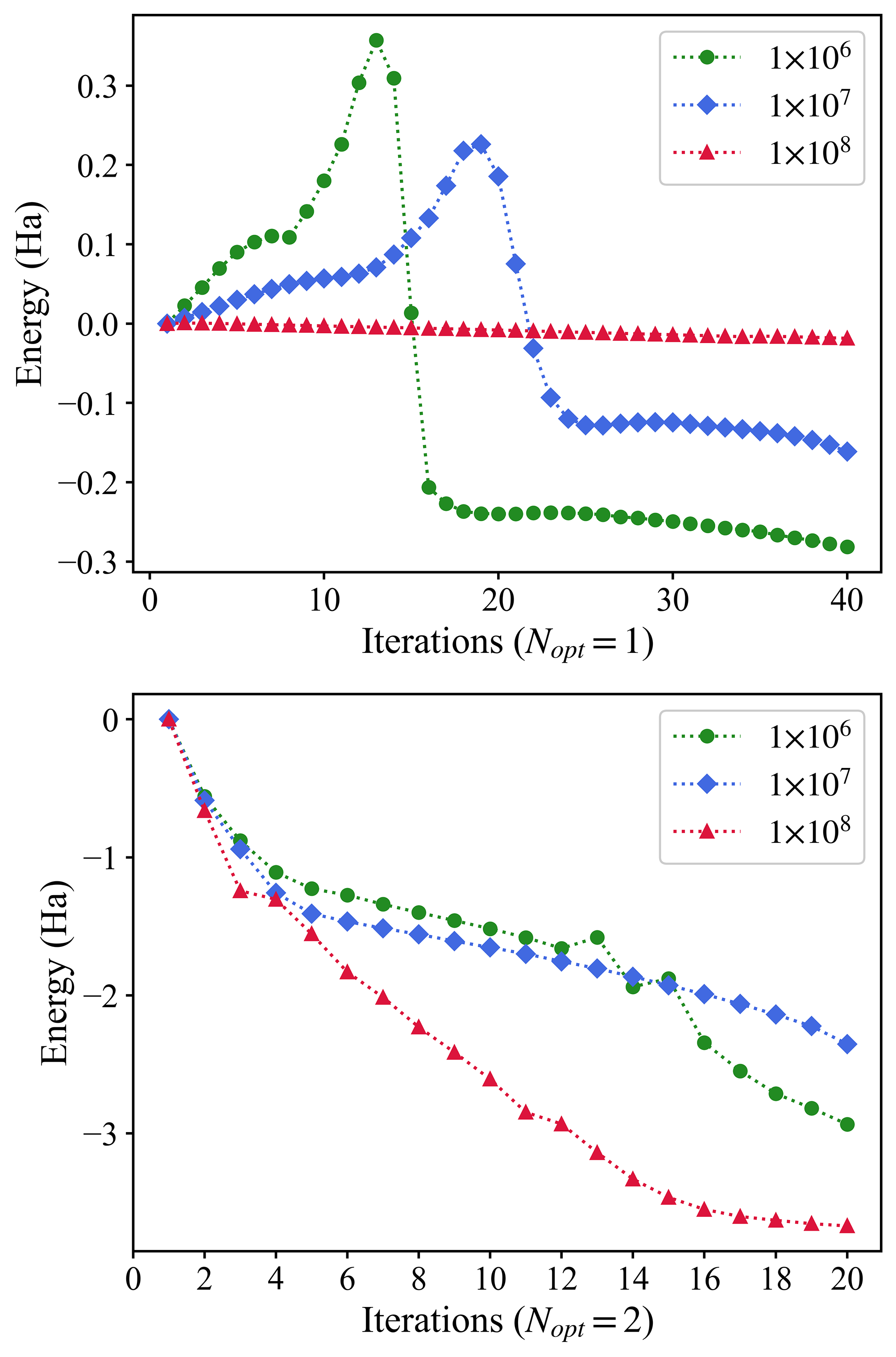}
    \caption{Performances of the constrained optimization scheme in computing the ground-state energy of $\rm N_2$ at a bond length of 2.8 Å using the cc-pVDZ basis set, focusing on two specific values for the number of network parameter updates per iteration ($N_{\rm opt}$). Additionally, three different sizes of Monte Carlo configuration sample are employed to evaluate NQS energy and its gradient: ${10}^6$, ${10}^7$, and ${10}^8$.}
    \label{fig:6}
\end{figure}

Another crucial insight from our analysis is the pronounced robustness of constrained optimization in larger molecular systems. For instance, Figure \ref{fig:6} contrasts the performances at $N_{\rm opt}=1$ and $N_{\rm opt}=2$ for $\rm N_2$ ground-state energy calculations in the cc-pVDZ basis set at a bond length of 2.8 Å. With $N_{\rm opt}=1$, focusing solely on standard VMC energy $ E_\theta$ leads to difficulties in reducing energy levels until the Monte Carlo sample size is substantially increased to ${10}^8$. Conversely, with $N_{\rm opt}=2$, optimizing both the standard VMC energy $E_\theta$ and the truncated energy ${\widetilde{E}}_\theta$ achieves rapid and consistent energy reduction. This is because the gradient of $E_\theta$, ${g}_\theta$, can only be approximated with the selected configuration sample, which may not effectively indicate the correct direction for parameter updates. However, within the constrained optimization framework, the gradient ${\widetilde{g}}_\theta$ of energy ${\widetilde{E}}_\theta$ can be precisely evaluated, allowing for a more direct and efficient approach to the minimum.

\subsection{\label{sec:perform_block}Performance of block optimization}
\begin{figure}[!h]
    \centering
    \includegraphics[width=0.45\textwidth]{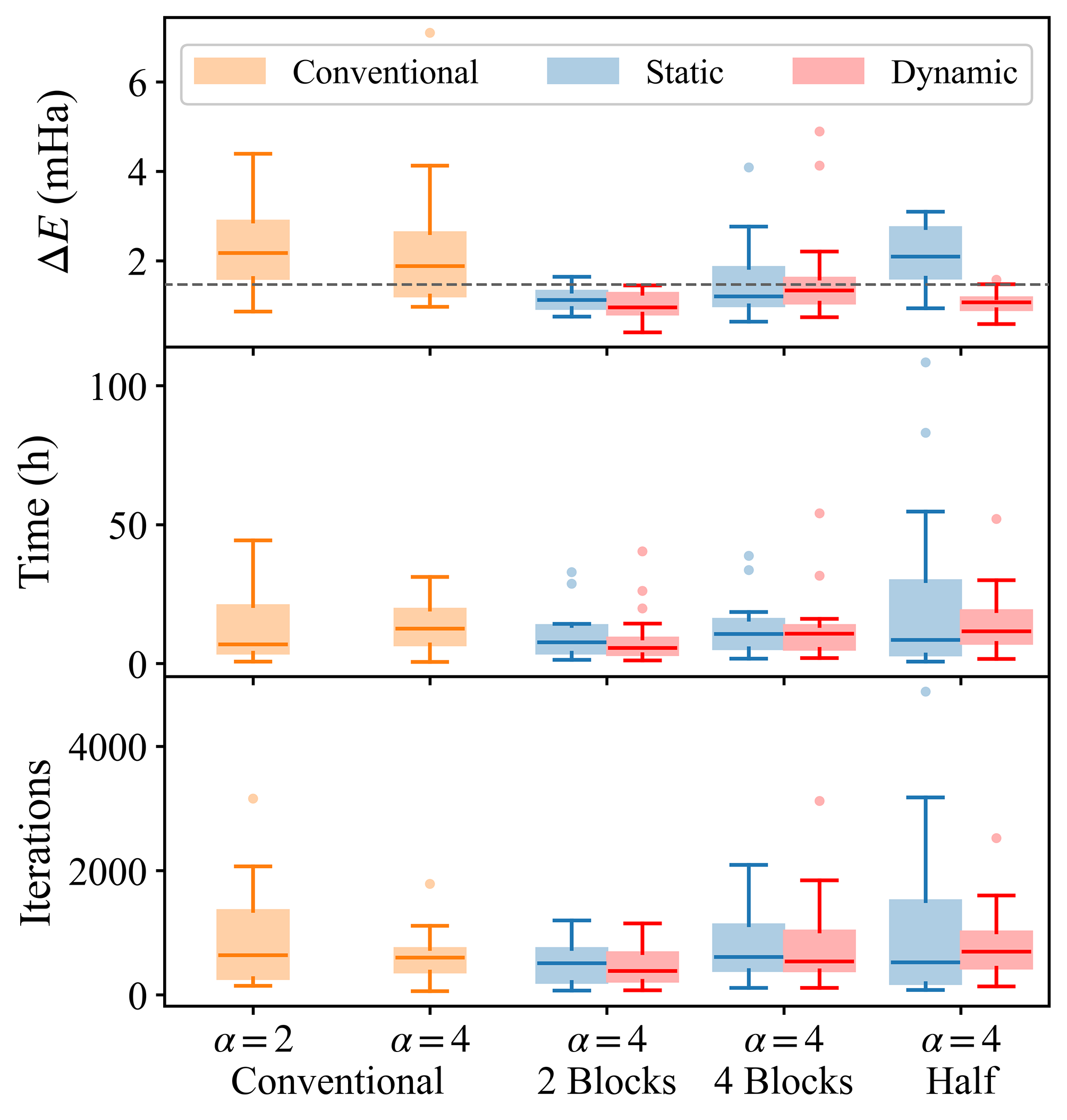}
    \caption{  Evaluations of the block optimization scheme applied to calculating the ground-state energies of $\rm H_2O$ across 20 O$-$H bond lengths within the 6-31G basis set. The top panel represents energy differences ($\Delta E$) relative to FCI results, with a gray line indicating the threshold of chemical accuracy. The middle and bottom panel show the wall time and the iterations necessary to approaching the same energy level (the highest converged energy across all strategies), respectively. The presentation of data follows the boxplot format established in Fig. \ref{fig:5}.}
    \label{fig:7}
\end{figure}

This section delves into the efficacy of block optimization in calculating the ground-state energies of $\rm H_2O$ in the 6-31G basis set, with the hidden unit density ($\alpha$) set to 4. In all scenarios, the adaptive learning rate algorithm facilitates optimization. We investigate the segmentation of the original RBM model into 2 and 4 blocks, employing both static and dynamic strategies for the division. The static strategy maintains a constant division of the RBM model throughout the optimization process, whereas the dynamic strategy involves a random re-division of the model at each parameter update. Additionally, we explore adapted static and dynamic 2-block schemes that optimize only one set of the two blocked parameters, referred to as the `Half'    scheme in Figure \ref{fig:7}.

Results depicted at the top of Figure \ref{fig:7} illustrate that all variations of block optimization, with the exception of the scheme optimizing only a fixed half of the parameters, achieve higher accuracy than conventional optimization approaches, regardless of whether $\alpha$ is 2 or 4. Notably, a dynamic division into 2 blocks—regardless of optimizing half or all blocked parameters—consistently reaches chemical accuracy in nearly all instances. This demonstrates that the noise introduced by block optimization in Eq. (\ref{eq:SR}) can reduce the likelihood of the NQS getting stuck in local minima. Moreover, the 2-block optimization marginally outperforms the 4-block strategy, with both fixed and dynamic approaches yielding similar results. It's important to highlight that optimizing a fixed half of the parameters shows comparable performance to conventional optimization when $\alpha=2$, as both essentially optimize an RBM model with an equivalent parameter count.

The experiments reveal no significant time savings with block optimization, primarily due to the minor differences in time required to solve Eq. (\ref{eq:SR}) in the context of a model with a relatively modest number of network parameters (2834 in this case). However, conducting comparative experiments using larger RBM models presents a challenge due to the prohibitive computational demands associated with conventional optimization methods.

\subsection{\label{sec:h2o_n2}Dissociation curves for $\mathbf{H_2O}$ and $\mathbf{N_2}$}
In this study, we employ a comprehensive VMC approach combining adaptive learning rate, constrained optimization, and block optimization to investigate the dissociation curves of $\rm H_2O$ and $\rm N_2$ in the cc-pVDZ basis set. These curves, representing the continuous variation of electron correlations, serve as favored benchmarks for evaluating advancements in electronic structure methods.

Utilizing an RBM model with a hidden unit density of 6, which includes 14,160 and 16,588 network parameters for $\rm H_2O$ and $\rm N_2$, respectively, we face the challenge of directly solving the SR equation, Eq. (\ref{eq:SR}), due to the extensive number of parameters. To address this, we employ a block optimization strategy, dynamically partitioning the extensive RBM model into two segments for each parameter update and optimizing one segment at a time.

\begin{figure*}
    \centering
    \includegraphics[width=0.9\textwidth]{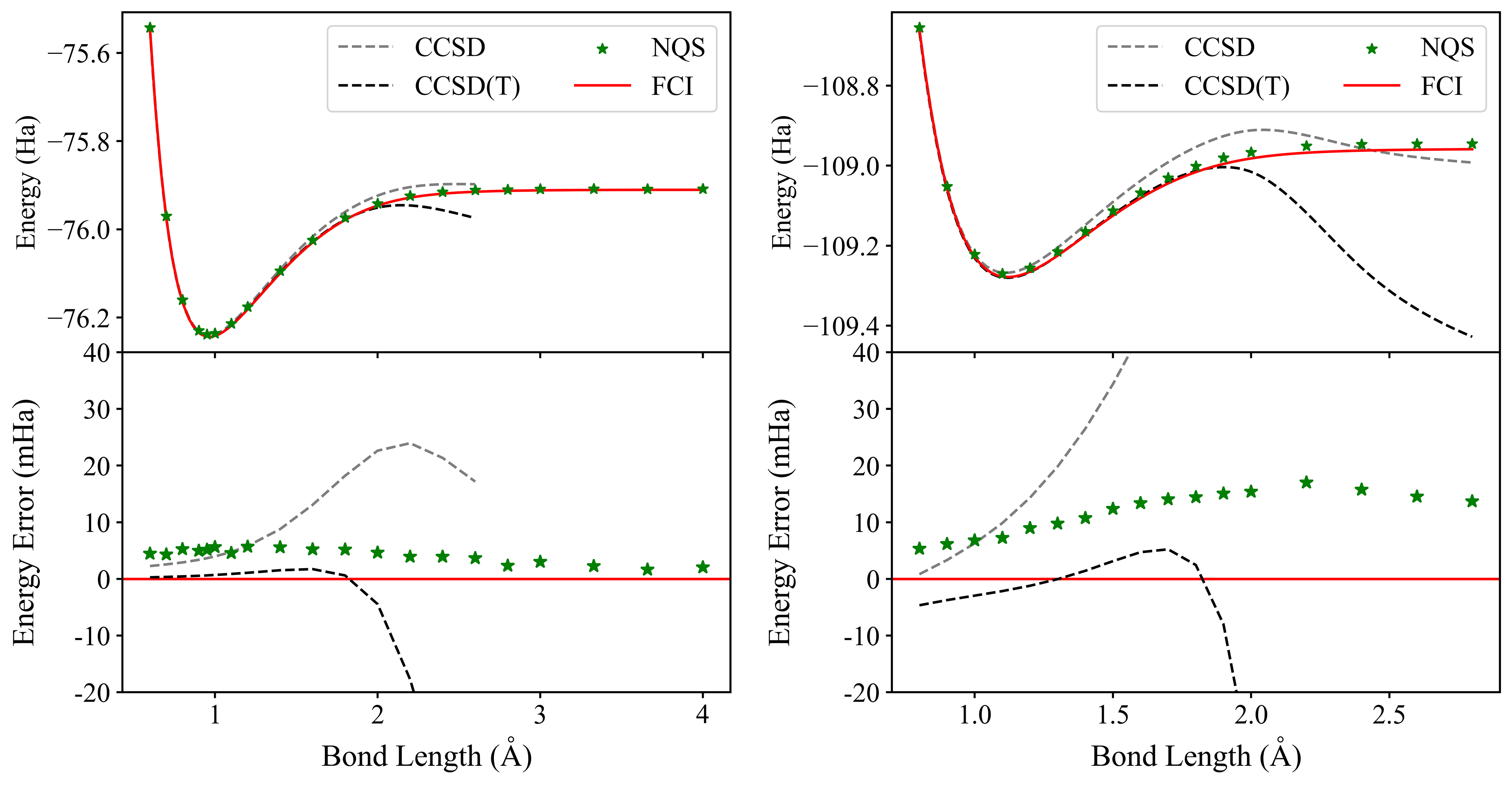}
    \caption{Dissociation curves (upper panel) and energy deviations relative to FCI results (lower panel) for both $\rm H_2O$ (left) and $\rm N_2$ (right) in the cc-pVDZ basis set.}
    \label{fig:8}
\end{figure*}

As shown in Figure \ref{fig:8}, this methodology accurately reproduces dissociation curves that closely track the FCI benchmark across the entire dissociation range for both molecules, highlighting the its robustness even where traditional CC methods fail due to the predominance of static correlation at extended bond lengths. The mean deviation from FCI results stands at approximately 4 millihartree (mHa) for $\rm H_2O$ and 12 mHa for $\rm N_2$, showcasing competitive or potentially superior performance relative to the Machine Learning Configuration Interaction (MLCI) method.\cite{MLCI19} Unlike the MLCI method, which relies on supervised learning to train neural networks and then calculates exact CI energy within a selected sample $\mathcal{V}$, our approach optimizes the NQS ansatz directly through the VMC framework, leveraging NQS's unique capacity to bootstrap and predict the coefficient function $\psi_\theta$. The observed discrepancies in energy are likely attributable to either the inherent limitations of the RBM model or optimization challenges, or possibly a combination of both factors.

\subsection{\label{sec:cr2}Ground-state energy of $\mathbf{Cr_2}$}
The chromium dimer, $\rm Cr_2$, poses a significant challenge in electronic structure theory due to its complex electronic structure and the unique nature of its potential energy curve.\cite{Cr2_22_jacs} Despite the limitations of the Ahlrichs SV basis set for providing an accurate description, the ground-state energy calculation of $\rm Cr_2$ at a bond length of 1.5 Å using this basis has become a popular benchmark for assessing the precision and computational efficiency of various electronic structure methods.\cite{Cr2_DMRG09,Cr2_DMRG15,FCIQMC14,ASCI16,iCI20,Cr2_HCI20} In this context, we evaluate the NQS's accuracy in both the active space CAS(24e,30o) and the full configuration space CAS(48e,42o) of this large, strongly correlated molecular system.

The optimization begins with a CASSCF(12e,12o) calculation, utilizing the derived natural molecular orbitals. The adaptive learning rate is implemented, dynamically choosing from 100 potential values within the range $\left[{10}^{-4},0.2\right]$. Constrained optimization, with $N_{\rm opt}=5$, is applied to ensure stable energy reduction. For block optimization, the RBM model, with a hidden unit density of $\alpha=12$, is dynamically partitioned into three blocks at each update, optimizing parameters in two of the three blocks. The input bias of the RBM model is initiated with a hyperparameter $T=1/6$, and the diagonal regularization parameter ($\lambda$) for the SR method is $10^{-5}$. To strike a balance between the accuracy and efficiency, the selected configuration sample $\mathcal{V}$ includes electron configurations with $\left|\psi_\theta\right|>4\times{10}^{-5}$ for NQS optimization.

As revealed in TABLE \ref{tab:cas_energy}, after approximately 70 optimization iterations, the NQS's variational energies in the active space already exceed the accuracy of the CCSDT method. The total energies, after perturbative correction (detailed in Appendix \ref{sec:PT}), outperform both the CCSD(T) and CCSDTQ methods. Moreover, initially selecting around 170,000 electron configurations with $\left|\psi_\theta\right|>1\times{10}^{-4}$ using the optimized NQS ansatz, the resultant variational energy $\widetilde{E}_\theta$ differs from the HCI energy with around 120,000 electron configurations by approximately 4mHa, with a perturbatively corrected energy discrepancy of about 3mHa. This suggests a high quality of the current truncated NQS ansatz close to the HCI wave function. A subsequent selection of roughly 390,000 electron configurations with $\left|\psi_\theta\right|>4\times{10}^{-5}$ yields a minor decrease (less than 4mHa) in variational energy, contrasting with a significant drop (about 15mHa) in the variational HCI energy with about 480,000 configurations. This indicates that enhancing NQS accuracy requires a much larger configuration sample $\mathcal{V}$ during optimization.

\begin{table}[!h]
\caption{\label{tab:cas_energy}Ground-state energies of the $\rm Cr_2$ in the active space CAS(24e,30o) obtained from different methods.}
\begin{ruledtabular}
\begin{tabular}{lccl}
\multicolumn{2}{c}{ }&\multicolumn{2}{c}{Energy(+2086Ha)}\\
Method&\makecell{Number of\\Configurations}&\makecell{Variational\\Energy}&\makecell{Total\\Energy}\\
\hline
CCSDT\cite{Cr2_DMRG09}   &-          &-          &$-0.38049$\\
CCSD(T)\cite{Cr2_DMRG15} &-          &-          &$-0.39864$\\
CCSDTQ\cite{Cr2_DMRG09,Cr2_DMRG15}  &-          &-          &$-0.40670$\\
\textbf{NQS}     &\textbf{174 963}    &$\mathbf{-0.38396}$   &$\mathbf{-0.41692}$\\
\textbf{NQS}    &\textbf{385 554}    &$\mathbf{-0.38752}$  &$\mathbf{-0.41631}$\\
HCI\cite{Cr2_HCI20}     &123 144    &$-0.38766$   &$-0.42009$\\
HCI\cite{Cr2_HCI20}     &480 138    &$-0.40286$   &$-0.42054$\\
HCI\cite{Cr2_HCI20}     &66 679 956 &$-0.42011$   &$-0.42138$\\
iCIPT2\cite{iCI20}  &-          &$-0.41613$   &$-0.42147$\\
ASCI\cite{ASCI16}    &1 000 000  &$-0.40388$   &$-0.4203$ \\
DMRG\cite{Cr2_DMRG09}    &-          &$-0.42052$   &$-0.42116$\\
FCIQMC\cite{FCIQMC14}  &-          &-          &$-0.4212$\\
\end{tabular}
\end{ruledtabular}
\end{table}

\begin{table}[!h]
\caption{\label{tab:full_energy}Ground-state energies of the $\rm Cr_2$ in the full configuration space CAS(48e,42o) obtained from different methods.}
\begin{ruledtabular}
\begin{tabular}{lccl}
\multicolumn{2}{c}{ }&\multicolumn{2}{c}{Energy(+2086Ha)}\\
Method&\makecell{Number of\\Configurations}&\makecell{Variational\\Energy}&\makecell{Total\\Energy}\\
\hline
CCSD(T)\cite{Cr2_DMRG15} &-          &-          &$-0.42223$\\
CCSDTQ\cite{Cr2_DMRG15}  &-          &-          &$-0.43024$\\
\textbf{NQS}     &\textbf{156 594}    &$\mathbf{-0.38978}$   &$\mathbf{-0.43753}$\\
\textbf{NQS}     &\textbf{331 397}    &$\mathbf{-0.39203}$  &$\mathbf{-0.43452}$\\
HCI\cite{Cr2_HCI20}     &190 937    &$-0.40500$  &$-0.44290$\\
HCI\cite{Cr2_HCI20}     &787 919    &$-0.42216$  &$-0.44346$\\
HCI\cite{Cr2_HCI20}     &148 589 206&$-0.44277$   &$-0.44456$\\
iCIPT2\cite{iCI20}  &-          &$-0.43505$   &$-0.44474$\\
ASCI\cite{ASCI16}    &-          &-          &$-0.44325$\\
DMRG\cite{Cr2_DMRG15}    &-          &$-0.44333$   &$-0.44447$\\
\end{tabular}
\end{ruledtabular}
\end{table}

TABLE \ref{tab:full_energy} shows that the variational energies in the full configuration space are marginally lower than those in the active space, highlighting the optimization challenges in this extensively correlated system. Nonetheless, after second-order perturbative corrections, the total energies obtained through the truncated NQS ansatz surpass the accuracy of two referenced CC methods, with a difference of about 7 mHa compared to other more accurate benchmarks, underscoring the potential of NQS despite the challenges.

\section{\label{sec:conclusion}Conclusion}

This study introduces a suite of algorithmic advancements aimed at enhancing the optimization of the NQS ansatz via orbital space VMC framework, encompassing adaptive learning rate, constrained optimization, and block optimization strategies. These improvements facilitate the application of the NQS ansatz to more complex molecular systems.

The adaptive learning rate algorithm, leveraging a truncated energy expression, incurs minimal additional computational cost---equivalent to roughly one extra energy evaluation---while achieving notable gains in both accuracy and efficiency. Constrained optimization, by reutilizing selected configurations and the truncated energy, along with employing exact energy gradients within the truncated subspace, facilitates fast and stable energy minimization. These methods, focusing on the intrinsic properties of NQS optimization, show potential for generalization across various neural network models.

Moreover, we have proposed a block optimization approach tailored for extensive RBM models, capitalizing on their unique analytical properties to reduce optimization complexity. By integrating these three algorithmic improvements, together with a parameter initialization strategy, we have successfully modeled the dissociation curves for $\rm H_2O$ and $\rm N_2$ in the cc-pVDZ basis set, as well as determined the ground-state energy of the strongly correlated $\rm Cr_2$ in the Ahlrichs SV basis.

Despite the progress, the expressive limitations of RBMs in capturing the complexities of larger molecules with strong electron correlations or intricate structures remain a challenge. Future efforts may explore more advanced deep neural-network architectures,\cite{NQS24-backflow,QianKunNet23} which promise enhanced accuracy albeit at higher computational costs. Therefore, we hope that the methodologies developed herein will facilitate more efficient and robust NQS optimization, broadening the scope of NQS applications in quantum chemistry.

\begin{acknowledgments}
This work was financially supported by the National Natural Science Foundation of China (Grant Nos. 22222605 and 22076095). H.-E. Li was also financially supported by the National Natural Science Foundation of China (Grant Nos. 223B1011). The Tsinghua Xuetang Talents Program and High-Performance Computing Center of Tsinghua University were acknowledged for providing computational resources.
\end{acknowledgments}
\section*{Author Declarations}
\subsection*{Conflict of Interest}
The authors have no conflicts to disclose.
\subsection*{Author Contributions}
X. Li, J.-C. Huang and G.-Z. Zhang contributed equally to this work.

\textbf{Xiang Li:} Conceptualization (equal); Data Curation (equal); Formal Analysis (equal); Investigation (equal); Methodology (equal); Project Administration (equal); Software (equal); Supervision (equal); Validation (equal); Visualization (equal); Writing – original draft (equal); Writing – review \& editing (equal)
\textbf{Jia-Cheng Huang:} Conceptualization (equal); Data Curation (equal); Formal Analysis (equal); Investigation (equal); Methodology (equal); Software (equal); Validation (equal); Visualization (equal); Writing – review \& editing (equal)
\textbf{Guang-Ze Zhang:} Conceptualization (equal); Data Curation (equal); Formal Analysis (equal); Investigation (equal); Methodology (equal); Software (equal); Validation (equal); Visualization (equal); Writing – review \& editing (equal)
\textbf{Hao-En Li:} Validation (supporting); Visualization (supporting); Writing – review \& editing (equal)
\textbf{Zhu-Ping Shen:} Validation (supporting); Visualization (supporting); Writing – review \& editing (equal)
\textbf{Chen Zhao:} Visualization (supporting); Writing – review \& editing (supporting)
\textbf{Jun Li:} Funding Acquisition (supporting); Project Administration (supporting); Supervision (supporting); Writing – review \& editing (supporting)
\textbf{Han-Shi Hu:} Funding Acquisition (equal); Project Administration (equal); Resources (equal); Supervision (equal); Writing – review \& editing (equal)

\section*{Data Availability Statement}
The data that support the findings of this study are available from the corresponding author upon reasonable request.

\appendix

\section{\label{sec:MO_Initialization}Network parameter initialization with molecular orbital energy}

We focus on the input biases $a_i$ of the RBM, which are uniquely associated with the spin orbitals. If a specific spin orbital is occupied within an electron configuration, denoted as $\sigma_i^k=1$ in Eq. (\ref{eq:RBM_psi}), then the probability $P(D_k)$ of this configuration includes the term $e^{a_i + a_i^*}$, representing the probability contribution of input bias $a_i$. Therefore, we investigate the impact of parameter $a_i$ during NQS optimization by evaluating $e^{a_i + a_i^*}$ before and after calculating the ground-state energy of $\rm H_2O$ with an O$-$H bond length of 1.0Å, using the 6-31G basis set.

\begin{figure}[!h]
    \centering
    \includegraphics[width=0.45\textwidth]{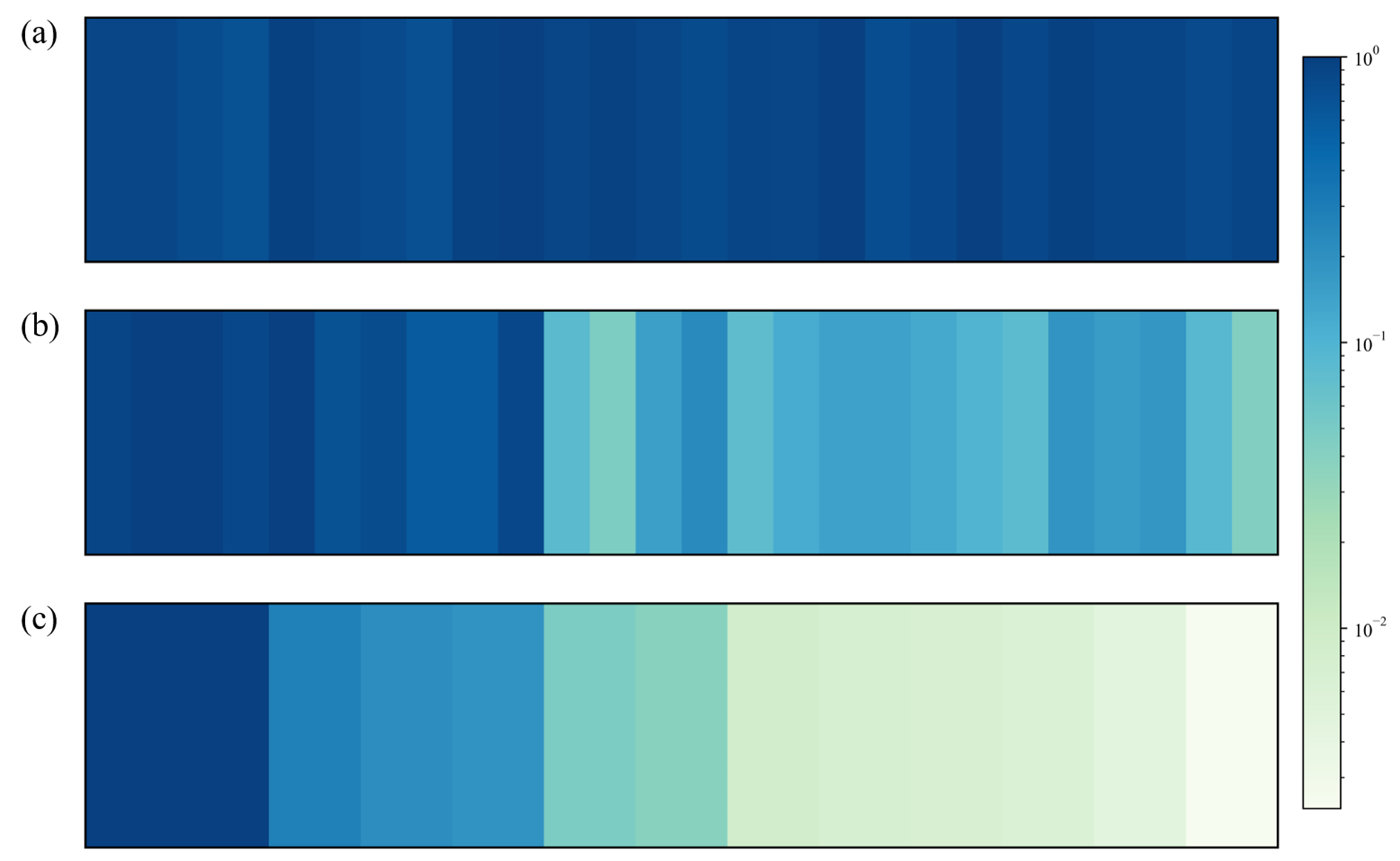}
    \caption{Graphic representation for the probability contribution of input biases $a_i$, in which the horizontal axis denotes the corresponding spin orbitals $\epsilon_i$ and is broadened along the vertical direction for clarity. The orbital energy increases from left to right. (a) The randomly initialized input biases before optimization. (b) The randomly initialized input biases after optimization. (c) The input biases initialized using orbital energy with $T=1$ in Eq. (\ref{eq:Boltzmann-Distribution}).}
    \label{fig:MO_init}
\end{figure}

As depicted in Figure \ref{fig:MO_init}, when the parameters are randomly initialized before NQS optimization, the probability contributions of input biases $a_i$ are similar, indicating that there is little difference in the probability of electron configurations regardless of which spin orbital the electron occupies. However, a distinctive distribution emerges after optimization, where the occupied spin orbitals with lower energy levels exhibit larger probability contributions, consistent with the fact that electrons prefer to occupy orbitals with lower energy levels.

This observation inspired a network parameter initialization strategy using the energy $\epsilon_i$ of molecular orbitals, which employs the Boltzmann distribution in Eq. (\ref{eq:Boltzmann-Distribution}) to approximate the optimized probability contribution of each input bias $a_i$. As shown in Figure \ref{fig:MO_init}, this simple initialization method yields a much better approximation than random initialization.

\section{\label{sec:PT}Derivation of the perturbative correction for NQS}

We derive the second-order perturbative correction formula for the energy of NQS based on the perturbation theory and Sharma's work.\cite{PT18} We first write the unperturbed ket $|\Psi_0\rangle$ and zero-order energy $E_0$ as
\begin{equation}
|\Psi_0\rangle = \sum_{k\in\mathcal{V}}c_k|D_k\rangle,
\end{equation}
\begin{equation}
E_0 = \langle \Psi_0|\hat{H}|\Psi_0\rangle = \sum_{k,t\in \mathcal{V}}c_k^*c_t \langle D_k|\hat{H}|D_t\rangle,
\end{equation}
where the configuration coefficients are
\begin{equation}
c_k=\frac{\psi_\theta(\sigma_1^k, \sigma_2^k, \ldots, \sigma_M^k)}{\sqrt{\sum_{t \in \mathcal{V}}|\psi_\theta(\sigma_1^t, \sigma_2^t, \ldots, \sigma_M^t)|^2}}.
\end{equation}

According to the perturbation theory, $|\Psi_0\rangle$ is an energy eigenket of the unperturbed Hamiltonian $\hat{H}_0$. For non-linearly parameterized wave function, the $\hat{H}_0$ can be defined as
\begin{equation}
\hat{H}_0  =\hat{P} \hat{H} \hat{P}+\hat{Q} \hat{H}^{\rm EN} \hat{Q},
\end{equation}
 where $\hat{P}=|\Psi_0\rangle\langle\Psi_0|$, and $\hat{Q}=\mathbf{1}-\hat{P}$. $\hat{H}^{\rm EN}$ is the Epstein-Nesbet Hamiltonian\cite{ENPT76}, defined as
 \begin{equation}
 \hat{H}^{\rm EN} = \sum_{k} E_k^{\rm HF} |D_k\rangle\langle D_k|,
 \end{equation}
 where $E_k^{\rm HF} = \langle D_k|\hat{H}|D_k\rangle$. This decomposition of $\hat{H}_0$ ensures that the zeroth order wave function and all the determinants that have a zero overlap with $\ket{\Psi_0}$ are eigenkets of $\hat{H}_0$. The perturbation is then given by $\hat{V}=\hat{H}-\hat{H}_0$. Introducing a shorthand  $\hat{R}\equiv (E_0-\hat{H}_0)^{-1}$, some useful properties of these operators are summarized below: 
\begin{enumerate}
\item The first order correction to the
wave function is \begin{equation}
|\Psi_1\rangle=\hat{Q} \hat{R} \hat{Q} \hat{V}|\Psi_0\rangle,
\end{equation} and \begin{equation}
\begin{aligned}
\hat{Q} \hat{V}|\Psi_0\rangle&=  (\hat{H}-\hat{H}_0)|\Psi_0\rangle-\hat{P}(\hat{H}-\hat{H}_0)|\Psi_0\rangle \\
& =(\hat{H}-E_0)|\Psi_0\rangle
\end{aligned}
\end{equation}
\item The approximation\begin{equation}
\langle D_k|\hat{R}| D_t\rangle \approx\langle D_k|\frac{1}{E_0-\hat{H}^{\rm EN}}| D_t\rangle=\frac{\delta_{k t}}{E_0-E_{k}^{\rm HF}}.
\end{equation}
\end{enumerate}

The second order correction $\Delta E_2$ to the wave function $|\Psi_0\rangle$ with energy $E_0$ is then given by
\begin{equation}
\begin{aligned}
\Delta E_2 & =\langle\Psi_0|\hat{V}| \Psi_1\rangle \\
& =\langle\Psi_0|(\hat{H}-E_0) \hat{R}(\hat{H}-E_0)| \Psi_0\rangle \\
& =\langle\hat{H} \hat{R} \hat{H}\rangle_{\Psi_0}-E_0\langle\{\hat{H},\hat{R}\}\rangle_{\Psi_0}+E_0^2\langle\hat{R}\rangle_{\Psi_0}.
\end{aligned}
\end{equation}
Taking advantage of the completeness $\sum_k |D_k\rangle\langle D_k|=\mathbf{1}$ and using the properties mentioned above, we can rewrite these terms as

\begin{equation}
\begin{aligned}
  \langle\hat{H} \hat{R} \hat{H}\rangle_{\Psi_0}&=\sum_{k t} \langle\Psi_0| \hat{H}| D_k \rangle \langle D_k| \hat{R}| D_t \rangle \langle D_t| \hat{H}| \Psi_0 \rangle \\
& =\sum_k \frac{ | \langle D_k| \hat{H}| \Psi_0 \rangle |^2}{E_0-E_{k}^{\rm HF}} ;
\end{aligned}
\end{equation}
\begin{equation}
\begin{aligned}
\langle\{\hat{H},\hat{R}\}\rangle_{\Psi_0}&=2  \operatorname{Re} \sum_{k t} \langle\Psi_0 
|D_k \rangle \langle D_k|\hat{R}| D_t \rangle \langle D_t| \hat{H}| \Psi_0 \rangle  \\
& =\sum_{k \in \mathcal{V}} \frac{2   \operatorname{Re} \{ \langle\Psi_0 | D_k \rangle \langle D_k| \hat{H}| \Psi_0 \rangle \}}{E_0-E_{k}^{\rm HF}} ;
\end{aligned}
\end{equation}
\begin{equation}
\begin{aligned}
 \langle\hat{R}\rangle_{\Psi_0}&=  \sum_{k t} \langle\Psi_0 | D_k \rangle \langle D_k|\hat{R}| D_t \rangle \langle D_t | \Psi_0 \rangle  \\
& =\sum_{k \in \mathcal{V}} \frac{  | \langle D_k | \Psi_0 \rangle |^2}{E_0-E_{k}^{\rm HF}} .
\end{aligned}
\end{equation}
Therefore, the resulting correction is
\begin{equation}
\begin{aligned}
\Delta E_2 & =\sum_{k \in \mathcal{V}}\frac{|\langle D_k| \hat{H}| \Psi_0 \rangle-c_k\times E_0|^2}{E_0-E_{k}^{\rm HF}}+\sum_{k \notin \mathcal{V}}\frac{| \langle D_k| \hat{H}| \Psi_0 \rangle |^2}{E_0-E_{k}^{\rm HF}}.
\end{aligned}
\end{equation}
In practice, the term $k\notin \mathcal{V}$ in the second term of $\Delta E_2$ is approximated by configurations selected from the single and double excitation space of sample $\mathcal{V}$.
\nocite{*}
\section*{References}
\bibliography{aipsamp}

\end{document}